# Relative and Mean Motions of Multi-Machine Power Systems in Classical Model

Bin Wang, *Student Member, IEEE*, Kai Sun, *Senior Member, IEEE*, and Wei Kang, *Fellow, IEEE*

*Abstract*—It is well-known that in an *m*-machine power system where each machine is represented by a second-order differential equation, the Jacobian of the system equation contains (*m*-1) pairs of conjugate eigenvalues and two real eigenvalues, including at least one zero. This letter proves that under the uniform damping condition, the dynamics associated with the two real eigenvalues do not have any impact on the dynamics associated with those complex eigenvalues. This conclusion is important to justify the use of the relative motions or center-of-inertia (COI) coordinate to analyze the rotor angle stability in a multi-machine power system.

*Index Terms*—Relative motion, mean motion, power system, rotor angle stability, synchronism, center-of-inertia (COI) coordinate.

## I. Introduction

ELECTRIC power system is always operated around a synchronous frequency/speed, i.e. 50Hz/60Hz. Power system stability is the ability of an electric power system, for a given initial operating condition, to regain a state of operating equilibrium after being subjected to a physical disturbances, with most system variables bounded so that practically the entire system remains intact [1]. Specifically, the rotor angle stability of the system refers to the ability of synchronous machines of an interconnected power system to remain in synchronism after being subjected to a disturbance [1]. The synchronism here is actually determined by the relative motions among the machines, this is the reason why the center-of-inertia (COI) coordinate is always suggested for studying the rotor angle stability.

This letter provides mathematical proof to justify why the use of the relative motions are sufficient to determine the rotor angle stability of a power system in classical model under the uniform damping condition.

## II. Relative and Mean Motions of Power Systems in Classical Model

Consider a general *m*-machine power system below [2].

$$\ddot{\delta}_i + \frac{D_i}{2H_i}\dot{\delta}_i + \frac{\omega_s}{2H_i}(P_{mi} - P_{ei}) = 0 \quad (1)$$

$$P_{ei} = E_i^2 G_i + \sum_{j=1, j \neq i}^{m} \left( C_{ij} \sin(\delta_i - \delta_j) + D_{ij} \cos(\delta_i - \delta_j) \right) \quad (2)$$

where $i \in \{1,2,...,m\}$, $\delta_i$, $P_{mi}$, $P_{ei}$, $E_i$, $H_i$ and $D_i$ respectively represent the absolute rotor angle, mechanical power, electrical power, electromotive force, the inertia constant and damping constant of machine $i$, and $G_i$, $C_{ij}$, and $D_{ij}$ represent network parameters including loads modeled by constant impedances.

Suppose that the system (1) has an asymptotically stable equilibrium point. It is well-known that the *m*-machine power system modeled by (1) has (*m*-1) pairs of conjugate complex eigenvalues, which respectively correspond to (*m*-1) oscillatory modes and represent the *relative motions* of the system, and two real eigenvalues (including one zero eigenvalue) which represent the *mean motion* of the system [3]. To define the relative/mean motion, a few notations are presented first.

Denote $\Delta = [\boldsymbol{\delta}^T \ \dot{\boldsymbol{\delta}}^T]^T$ as the state vector and rewrite (1) into the first-order representation as (3), where $\boldsymbol{\delta} = [\delta_1, \delta_2,..., \delta_m]^T$. Let its asymptotically stable equilibrium point be $\boldsymbol{\delta}_s = [\delta_{s1}, \delta_{s2},..., \delta_{sm}]^T$. Then, calculate its Jacobian matrix $J$ at $\boldsymbol{\delta} = \boldsymbol{\delta}_s$ in (4) [4]. Without loss of generality, denote $J$'s (*m*-1) pairs of conjugate eigenvalues as $\lambda_{2i-1}$ and $\lambda_{2i}$ ($i = 1, 2,..., m$-1) and its two real eigenvalues as $\lambda_{2m-1}$ and $\lambda_{2m}$. Let $R$ and $L$ be the matrices whose columns and rows are respectively $J$'s right and left eigenvectors. Assume $R = L^{-1}$. Using matrix $R$ or $L$ as the linear coordinate transformation as shown in (9), we obtain the system model in the modal space as shown in (10), where $Y = [y_1, y_2,..., y_{2m}]^T$. Note that $y_i$ is associated with $\lambda_i$ such that $y_{2m-1}$ and $y_{2m}$ define the *mean motion* while all other $y_i$ for $i=1, 2, ..., 2m$-2 define the *relative motions*.

$$\dot{\Delta} = \mathbf{f}(\Delta) \quad (3)$$

$$J = \begin{bmatrix} \mathbf{0} & I \\ N & M \end{bmatrix} \quad (4)$$

where $\mathbf{0}$ is an $m \times m$ zero matrix, $I$ is an $m \times m$ identity matrix, $M$ and $N$ are $m \times m$ diagonal matrices shown below. Assume that the system has a uniform damping, i.e. $D_i/2H_i = D_j/2H_j = c$ for any $i, j$, where $c$ is a constant.

$$\begin{cases} [M]_{ii} = D_i/2H_i = c \\ [M]_{ij} = 0 \end{cases} \Rightarrow M = cI \quad (5)$$

This work was supported by NSF CAREER Award (ECCS-1553863).
B. Wang and K. Sun are with University of Tennessee, Knoxville, TN 37996 USA. (e-mail: {bwang, kaisun}@utk.edu).
W. Kang is with the Department of Applied Mathematics, Naval Postgraduate School, Monterey, CA 93943 USA. (e-mail: wkang@nps.edu).

$$\begin{cases} [N]_{ij} = \frac{\omega_s}{2H_i}\left(C_{ij}\cos(\delta_{si}-\delta_{sj}) - D_{ij}\sin(\delta_{si}-\delta_{sj})\right) \\ [N]_{ii} = -\sum_{j=1,j\neq i}^{m}[N]_{ij} \end{cases} \quad (6)$$

$$R = \begin{bmatrix} \mathbf{r}_1 & \mathbf{r}_2 & \cdots & \mathbf{r}_{2m} \end{bmatrix} \quad (7)$$

$$L = \begin{bmatrix} \mathbf{l}_1 & \mathbf{l}_2 & \cdots & \mathbf{l}_{2m} \end{bmatrix}^T \quad (8)$$

$$\Delta = RY \text{ or } Y = L\Delta \quad (9)$$

$$\dot{Y} = R^{-1}\mathbf{f}(RY) = L\mathbf{f}(RY) \quad (10)$$

III. UNDERSTANDING THE RELATIVE AND MEAN MOTIONS

This section proposes the following two claims and presents their proofs, where all involved assumptions are first restated.

**Asm. 1: All machines are represented by $2^{nd}$-order model and all loads are represented by constant impedance.**
**Asm. 2: The steady-state condition of the power system is an asymptotically stable equilibrium point.**
**Asm. 3: The eigenvalues of the Jacobian of the $m$-machine power system contains ($m$-1) pairs of complex conjugate.**
**Asm. 4: The power system has uniform damping.**

*Claim 1*: For the power system in (1)-(10), its mean motion does not have any impact on its relative motions.
*Claim 2*: For the power system in (1)-(10), its relative motions can be represented by an ($m$-1)-oscillator system.
(An $N$-oscillator system is a dynamical system whose Jacobian has all its eigenvalues with non-zero imaginary parts)

*Lemma 1.1*: The sum of any column of matrix $N$ is zero.
*Lemma 1.2*: Matrix $J$ has a zero eigenvalue and has another real eigenvalue equal to $c$. (Without loss of generality, let $\lambda_{2m-1} = 0$ and $\lambda_{2m} = c$).
*Lemma 1.3*: In the right eigenvector $\mathbf{r}_{2m-1} = [r_{1,2m-1}, r_{2,2m-1},\ldots, r_{2m,2m-1}]^T$, $r_{i,2m-1} = r_{j,2m-1}$ holds for any $i, j$ from $\{1, 2, \ldots, m\}$ and $r_{i,2m-1} = 0$ holds for any $i$ from $\{m+1, m+2, \ldots, 2m\}$.
*Lemma 1.4*: In the right eigenvector $\mathbf{r}_{2m} = [r_{1,2m}, r_{2,2m},\ldots, r_{2m,2m}]^T$, $r_{i,2m} = r_{j,2m}$ holds for any $i, j$ from $\{1, 2, \ldots, m\}$ and $r_{i,2m} = r_{j,2m}$ holds for any $i, j$ from $\{m+1, m+2, \ldots, 2m\}$.

*Proof of Lemma 1.1*: Eq. (6) directly proves this lemma. ∎

$$\sum_{j=1}^{m}[N]_{ij} = [N]_{ii} + \sum_{j=1,j\neq i}^{m}[N]_{ij} = -\sum_{j=1,j\neq i}^{m}[N]_{ij} + \sum_{j=1,j\neq i}^{m}[N]_{ij} = 0 \quad (11)$$

*Proof of Lemma 1.2*: By adding all rows of matrix $N$ except for row $i$ to its row $i$, a zero row vector will be obtained according to lemma 1.1. Thus, $\det(N) = 0$ and we have (12), which shows that matrix $J$ has a zero eigenvalue. Recall (5) and we have (13), which shows that $c$ is an eigenvalue of $J$. ∎

$$\det(J) = \det(\mathbf{0})\det(M) - \det(I)\det(N) = 0 - 0 = 0 \quad (12)$$

$$\det(J - cI) = \det(\begin{bmatrix} -cI & I \\ N & \mathbf{0} \end{bmatrix}) \\ = \det(-cI)\det(\mathbf{0}) - \det(I)\det(N) = 0 - 0 = 0 \quad (13)$$

*Proof of Lemma 1.3*: Denote $\mathbf{r}_{2m-1,\text{ang}} = [r_{1,2m-1}, r_{2,2m-1},\ldots, r_{m,2m-1}]^T$ and $\mathbf{r}_{2m-1,\text{spe}} = [r_{m+1,2m-1}, r_{m+2,2m-1},\ldots, r_{2m,2m-1}]^T$, then $\mathbf{r}_{2m-1} = [\mathbf{r}^T_{2m-1,\text{ang}}, \mathbf{r}^T_{2m-1,\text{spe}}]^T$. Based on the definition of eigenvalue and right eigenvector and recall that $\lambda_{2m-1} = 0$, we obtain (14) and (15). The first equation in (15) shows that $r_{i,2m-1} = 0$ holds for any $i$ from $\{m+1, m+2, \ldots, 2m\}$. The second equation of (15) along with Lemma 1.1 shows that $r_{i,2m-1} = r_{j,2m-1}$ holds for any $i, j$ from $\{1, 2, \ldots, m\}$. ∎

$$J\mathbf{r}_{2m-1} = \lambda_{2m-1}\mathbf{r}_{2m-1} = 0 \Rightarrow \begin{bmatrix} \mathbf{0} & I \\ N & M \end{bmatrix}\begin{bmatrix} \mathbf{r}_{2m-1,\text{ang}} \\ \mathbf{r}_{2m-1,\text{spe}} \end{bmatrix} = 0 \quad (14)$$

$$\begin{cases} \mathbf{r}_{2m-1,\text{spe}} = 0 \\ N\mathbf{r}_{2m-1,\text{ang}} + M\mathbf{r}_{2m-1,\text{spe}} = N\mathbf{r}_{2m-1,\text{ang}} = 0 \end{cases} \quad (15)$$

*Proof of Lemma 1.4*: Denote $\mathbf{r}_{2m,\text{ang}} = [r_{1,2m}, r_{2,2m},\ldots, r_{m,2m}]^T$ and $\mathbf{r}_{2m,\text{spe}} = [r_{m+1,2m}, r_{m+2,2m},\ldots, r_{2m,2m}]^T$, then $\mathbf{r}_{2m} = [\mathbf{r}^T_{2m,\text{ang}}, \mathbf{r}^T_{2m,\text{spe}}]^T$. Similarly, we obtain (16). Recall (5), the second equation of (16) becomes $N\mathbf{r}_{2m,\text{ang}} = \mathbf{0}$. Similar to (15) in lemma 1.3, we have $r_{i,2m} = r_{j,2m}$ holds for any $i, j$ from $\{1, 2, \ldots, m\}$. Thus, the first equation of (16) shows that $r_{i,2m} = r_{j,2m} = c\, r_{1,2m}$ holds for any $i, j$ from $\{m+1, m+2, \ldots, 2m\}$. ∎

$$\begin{cases} \mathbf{r}_{2m,\text{spe}} = c\mathbf{r}_{2m,\text{ang}} \\ N\mathbf{r}_{2m,\text{ang}} + M\mathbf{r}_{2m,\text{spe}} = c\mathbf{r}_{2m,\text{spe}} \end{cases} \quad (16)$$

*Proof of Claim 1*: We only need to show that any relative motion, i.e. $\delta_i - \delta_j$, is independent of the mean motion, i.e. $y_{2m-1}$ and $y_{2m}$. Let $r_{a,b}$ be the element of matrix $R$ on row $a$ and column $b$. Then, (17) is obtained from (9). Since $r_{i,2m} = r_{j,2m}$ and $r_{i,2m-1} = r_{j,2m-1}$ for any $i, j$ by lemmas 1.3 and 1.4, then we have (18) and finish the proof. ∎

$$\delta_i - \delta_j = \sum_{k=1}^{2m}(r_{i,k} - r_{j,k})y_k = \sum_{k=1}^{2m-2}(r_{i,k} - r_{j,k})y_k \\ + (r_{i,2m} - r_{j,2m})y_{2m} + (r_{i,2m-1} - r_{j,2m-1})y_{2m-1} \quad (17)$$

$$\delta_i - \delta_j = \sum_{k=1}^{2m-2}(r_{i,k} - r_{j,k})y_k \quad (18)$$

*Lemma 2.1*: **In the left eigenvector $\mathbf{l}_i$ corresponding to a certain complex eigenvalue $\lambda_i$ as shown in (19), i.e. $i$=1,2, … or $2m$-2, the sum of the first $m$ elements is zero, i.e. (20) holds.**

$$\mathbf{l}_i = \begin{bmatrix} l_{i,1} & \cdots & l_{i,m} & l_{i,m+1} & \cdots & l_{i,2m} \end{bmatrix}^T \quad (19)$$

$$\sum_{j=1}^{m} l_{i,j} = 0 \text{ and } \sum_{j=m+1}^{2m} l_{i,j} = 0 \quad (20)$$

*Lemma 2.2*: **For any $i = 1,2,\ldots, m$, the electrical power of machine $i$, i.e. $P_{ei}$ in (2), does not depend on the mean motion.**

*Proof of Lemma 2.1*: Denote $\mathbf{l}_{i,\text{ang}} = [l_{i,1}, l_{i,2},\ldots, l_{i,m}]^T$ and $\mathbf{l}_{i,\text{spe}} = [l_{i,m+1}, l_{i,m+2},\ldots, l_{i,2m}]^T$, then $\mathbf{l}_i = [\mathbf{l}^T_{i,\text{ang}}, \mathbf{l}^T_{i,\text{spe}}]^T$. Based on the definition of eigenvalue and left eigenvector, we obtain (21) and (22). Recall (5), eliminate $\mathbf{l}^T_{i,\text{spe}}$ and obtain (23). The two sides of (23) are actually two column vectors, their sums should be equal, i.e. sum(LHS) = sum(RHS). Note that sum(LHS) = 0 by lemma 1.1 as shown in (24). Then, we have (25). Since $\lambda_i$ is a complex eigenvalue which guarantees that $\lambda_i (\lambda_i - c) \neq 0$, then we have proved the first equation in (20). With the second eq. of (22) we can similarly proof the second eq. in (20). ∎

$$\mathbf{l}^T_i J = \lambda_i \mathbf{l}^T_i \Rightarrow \begin{bmatrix} \mathbf{l}^T_{i,\text{ang}} & \mathbf{l}^T_{i,\text{spe}} \end{bmatrix} \begin{bmatrix} 0 & I \\ N & M \end{bmatrix} = \lambda_i \begin{bmatrix} \mathbf{l}^T_{i,\text{ang}} & \mathbf{l}^T_{i,\text{spe}} \end{bmatrix} \quad (21)$$

$$\begin{cases} \mathbf{l}^T_{i,\text{spe}} N = \lambda_i \mathbf{l}^T_{i,\text{ang}} \\ \mathbf{l}^T_{i,\text{ang}} + \mathbf{l}^T_{i,\text{spe}} M = \lambda_i \mathbf{l}^T_{i,\text{spe}} \end{cases} \quad (22)$$

$$N^T \mathbf{l}_{i,\text{ang}} = \lambda_i (\lambda_i - c) \mathbf{l}_{i,\text{ang}} \quad (23)$$

$$\text{sum(LHS)} = \sum_{j=1}^{m} \sum_{k=1}^{m} [N]_{jk} l_{i,k} = \sum_{k=1}^{m} l_{i,k} \sum_{j=1}^{m} [N]_{jk} = 0 \quad (24)$$

$$\text{sum(RHS)} = \lambda_i (\lambda_i - c) \sum_{j=1}^{m} l_{i,j} = \text{sum(LHS)} = 0 \quad (25)$$

*Proof of Lemma* 2.2: This is obvious according to Claim 1. ∎

*Proof of claim* 2: We only need to show that $y_{2m-1}$ and $y_{2m}$ do not appear on any of the first ($2m$-2) elements on the right hand side of (12). First expand the expression of **f** as shown in (26). Then, for any $i = 1, 2, \ldots,$ or $2m$-2, the row $i$ of (12) is shown in (27). Next, we only need to show the first and second terms in (27) do not depend on $y_{2m-1}$ and $y_{2m}$. Recall (9), the first term in (27) becomes (28) which is independent of $y_{2m-1}$ and $y_{2m}$, where lemmas 1.3, 1.4 and 2.1 have been applied. Similarly, the second term in (27) also does not depend on $y_{2m-1}$ and $y_{2m}$. ∎

## IV. Discussion

When any of the four assumptions in section III is not satisfied, claims 1 and 2 will not hold, neither. Especially, if assumption 4 is not satisfied, then lemmas 1.2, 1.4 and 2.1 will no longer hold. Any violation of the four assumptions will lead to two facts that (i) the mean motion of the power system will affect its relative motions; (ii) an error will be inevitably introduced in the rotor angle stability analysis if the mean motion is totally ignored.

$$\mathbf{f}(\Delta) = \begin{bmatrix} \dot{\delta}_1 \\ \vdots \\ \dot{\delta}_m \\ -c\dot{\delta}_1 - \dfrac{\omega_s}{2H_1}(P_{m1} - P_{e1}) \\ \vdots \\ -c\dot{\delta}_m - \dfrac{\omega_s}{2H_m}(P_{mm} - P_{em}) \end{bmatrix} \quad (26)$$

$$\dot{y}_i = \sum_{j=1}^{m} \left( l_{i,j} \dot{\delta}_j + l_{i,m+j} \left( -c\dot{\delta}_j - \dfrac{\omega_s}{2H_j}(P_{mj} - P_{ej}) \right) \right)$$
$$= \sum_{j=1}^{m} l_{i,j} \dot{\delta}_j - \sum_{j=1}^{m} c l_{i,m+j} \dot{\delta}_j - \sum_{j=1}^{m} \dfrac{\omega_s}{2H_j} l_{i,m+j} (P_{mj} - P_{ej}) \quad (27)$$

$$\sum_{j=1}^{m} l_{i,j} \dot{\delta}_j = \sum_{j=1}^{m} l_{i,j} \sum_{k=1}^{2m} r_{j,k} \dot{y}_k = \sum_{k=1}^{2m} \dot{y}_k \sum_{j=1}^{m} l_{i,j} r_{j,k}$$
$$= \dot{y}_{2m} \sum_{j=1}^{m} l_{i,j} r_{j,2m} + \dot{y}_{2m-1} \sum_{j=1}^{m} l_{i,j} r_{j,2m-1} + \sum_{k=1}^{2m-2} \dot{y}_k \sum_{j=1}^{m} l_{i,j} r_{j,k}$$
$$= \dot{y}_{2m} r_{j,2m} \sum_{j=1}^{m} l_{i,j} + \dot{y}_{2m-1} r_{j,2m-1} \sum_{j=1}^{m} l_{i,j} + \sum_{k=1}^{2m-2} \dot{y}_k \sum_{j=1}^{m} l_{i,j} r_{j,k} \quad (28)$$
$$= \sum_{k=1}^{2m-2} \dot{y}_k \sum_{j=1}^{m} l_{i,j} r_{j,k}$$